\documentclass[12pt]{article}
\usepackage{amssymb,amsfonts}
\usepackage{amsmath}
\usepackage{epsf,epsfig}
\begin{document}
\baselineskip 18pt

\title{Thermodynamics of the XX spin chain in the QTM approach}
\author{P.~N.~Bibikov}

\maketitle

\vskip5mm

\begin{abstract}
The free energy density of the XX chain in a magnetic field is obtained in two alternative ways within the Quantum Transfer Matrix approach.
In both cases the calculations are complete and self-consistent. All the intermediate constructions are presented explicitly in detail.
\end{abstract}

\maketitle
\vskip20mm

\section{Introduction}

Based on the Algebraic Bethe Ansatz \cite{1} and the Trotter-Suzuki formula \cite{2,3}, the Quantum Transfer Matrix (QTM) method \cite{4,5,6,7} produces a powerful machinery for evaluation of various thermodynamical properties for integrable spin chains.
Up to now, it was mainly applied to the Ising-like (easy-axis) XXZ chain, the most popular spin model \cite{1}.

All the QTM results on the XX (extremely easy-plane) chain are usually presented as analytical continuations of the corresponding XXZ ones \cite{8,9}.
Namely, the XXZ Hamiltonian depends on the parameter $\Delta$. For easy-axis models one has $|\Delta|>0$, while for easy-plane ones it will be
$|\Delta|<0$. Traditionally, in the former case, one puts $|\Delta|=\cosh{\eta}$, while in the latter $|\Delta|=\cos{\eta}$. In both cases the resulting formulas
have more transparent forms just in terms of $\eta$ (not in terms of $\Delta$). Within this paradigm, it is naturally to suppose that the XX chain, related to $\Delta=0$, should be considered as the $\eta=\pi/2$ easy-plane chain. However, in \cite{8,9} it was treated as the $\eta=i\pi/2$
easy-axis model.

For confirmation of the results, obtained long ago by several alternative approaches \cite{10,11,12}, this strategy is rather reasonable. At the same time, the QTM treatment of the XXZ model has some gaps. Namely, some basic assumptions in \cite{4,5}, have been verified only by numerical calculations corresponding to finite small Trotter numbers. These gaps have been filled in \cite{6}, however, in the manner which does not
further the calculation machinery intuition. Also, reading \cite{4,5,6,7}, it is not easy to understand what constructions are inherent in the QTM
approach in itself, and what caused by the complexity of the XXZ model.

In the present paper, studying the extremely simple (but not yet trivial!) XX chain, we give explicitly the detailed, step by step calculation
of its free energy density, filling the gaps of \cite{4,5}. Hence, we suppose that our paper supplements \cite{6}. Due to the simplicity of the XX chain, the intermediate constructions in our paper are much simpler than their XXZ analogs. For our opinion, this helps to reveal the QTM calculation machinery in its pure form.

The outline of the paper is the following. In Sect. 2, basing on the Yang-Baxter equation, we express the free energy density
at zero magnetic field from the dominant (leading, maximal) eigenvalue of the QTM transfer matrix. Though, the content of this section has been already presented in \cite{4,5,6,7}, we give  our own presentation in order to provide self-consistence of the paper. In Sect. 3 we study the infinite-temperature case, and show how it may be elementary
treated in the manner of \cite{13}. In Sect. 4, using the Algebraic Bethe Ansatz in the QTM framework, we obtain the so called dominant eigenvalue and the corresponding dominant
eigenvector of the quantum transfer matrix. In Sect. 5 we give the complete description of the associated Bethe and hole-type roots. In Sect. 6, taking the limit ${\tt N}\rightarrow\infty$, we get the resulting expression \eqref{fh0}, using manipulations with contour integrals. In Sect. 7 we
alternatively duplicate this result, using manipulations with Fourier transformations. In Sect. 8 we briefly
discuss the modifications, arising under introduction of a magnetic field. Finally, in Sect. 9 we enumerate all the QTM
constructions whose explicit forms have been obtained for the first time just in the present paper (due to the simplicity of the XX model).
We also discuss the additional complications, which are not inherent in the QTM approach in itself, but appear specifically in the XXZ case.

\section{Foundations of the QTM approach}

The keystone of the QTM approach \cite{4,5,6,7} is a $m^2\times m^2$ $R$-matrix
which satisfies the Yang-Baxter equation \cite{1}
\begin{equation}\label{YB}
R_{12}(\lambda-\mu)R_{23}(\lambda)R_{12}(\mu)=R_{23}(\mu)R_{12}(\lambda)R_{23}(\lambda-\mu),
\end{equation}
and at the vicinity of $\lambda=0$ takes the form
\begin{equation}\label{rh}
R(\lambda)=I^{(m^2)}+{\cal C}\lambda H+o(\lambda).
\end{equation}
Here $R_{12}\equiv R\otimes I^{(m)}$ and $R_{23}\equiv I^{(m)}\otimes R$, where $I^{(m)}$ denotes the $m\times m$ identity matrix. The matrix $H$
has the sense of the local Hamiltonian density for the periodic Hamiltonian
\begin{equation}
\hat H=\sum_{n=1}^NH_{n,n+1},\qquad H_{N,N+1}\equiv H_{N,1}.
\end{equation}
The latter acts in the so called {\it quantum space}, which is the tensor product of $N$ {\it local quantum spaces} ${\mathbb C}^m$ attached to the sites of the chain. Each $H_{n,n+1}$ acts as $H$ on the tensor product of two neighboring local quantum spaces and as $I^{(m)}$ on the other
tensor factors. The auxiliary numerical factor $\cal C$ usually is taken only for convenience, and may be reduced to unity by renormalization of $\lambda$.

In our case $m=2$, ${\cal C}=2$, and
\begin{equation}\label{rmat}
R(\lambda)=\left(\begin{array}{cccc}
1&0&0&0\\
0&\frac{\displaystyle1}{\displaystyle\cos{\lambda}}&\tan{\lambda}&0\\
0&\tan{\lambda}&\frac{\displaystyle1}{\displaystyle\cos{\lambda}}&0\\
0&0&0&1
\end{array}\right).
\end{equation}
The corresponding $H$ (${\bf S}^{\pm}$ and ${\bf S}^z$ are the usual spin-1/2 operators)
\begin{equation}\label{hamdens}
H=\frac{1}{2}\Big({\bf S}^+\otimes{\bf S}^-+{\bf S}^-\otimes{\bf S}^+\Big),
\end{equation}
is the Hamiltonian density matrix for the XX chain \cite{10,11,12}.

Using the substitutions,
\begin{equation}\label{L}
R(\lambda)=PL(\lambda),\qquad R(\lambda)=\tilde L(\lambda)P,
\end{equation}
(since $[P,R(\lambda)]=0$, in fact $L(\lambda)=\tilde L(\lambda)=PR(\lambda)$) where
\begin{equation}
P=\left(\begin{array}{cccc}
1&0&0&0\\
0&0&1&0\\
0&1&0&0\\
0&0&0&1
\end{array}\right),
\end{equation}
is the permutation matrix in the space ${\mathbb C}^2\otimes{\mathbb C}^2$ ($P\xi\otimes\eta=\eta\otimes\xi$, $\xi,\eta\in{\mathbb C}^2$) one may rewrite \eqref{YB} (after rather elementary manipulations) in the two equivalent forms
\begin{eqnarray}\label{RLL1}
&&R_{12}(\lambda-\mu)L_{13}(\lambda)L_{23}(\mu)=L_{13}(\mu)L_{23}(\lambda)R_{12}(\lambda-\mu),\\\label{RLL2}
&&R_{12}(-\mu-(-\lambda))\tilde L_{23}(-\mu)\tilde L_{13}(-\lambda)=\tilde L_{23}(-\lambda)\tilde L_{13}(-\mu)R_{12}(-\mu-(-\lambda)).
\end{eqnarray}

Contrary to the usual Algebraic Bethe Ansatz framework \cite{1}, we treat the $4\times4$ matrices $L(\lambda)$
and $\tilde L(\lambda)$ in \eqref{L} (the so called $L$-operators) as $2\times2$ matrices in the
local quantum space, whose entries are $2\times2$ matrices in the so called {\rm auxiliary} space. Within this approach,
the, so called, monodromy matrices \cite{1}
\begin{equation}
\tilde T_a(\lambda)=\tilde L_{N,a}(\lambda)\dots\tilde L_{1,a}(\lambda),\qquad
T_a(\lambda)=L_{1,a}(\lambda)\dots L_{N,a}(\lambda),
\end{equation}
are the $2^N\times2^N$ matrices in the quantum space (the tensor product of $N$ local quantum spaces) whose entries are the $2\times2$ matrices in the {\rm auxiliary} space (common for all $L$-operators).
The corresponding transfer matrices
\begin{equation}\label{t=TrT}
\tilde t(\lambda)\equiv{\rm tr}_a\tilde T_a(\lambda),\qquad
t(\lambda)\equiv{\rm tr}_a T_a(\lambda),
\end{equation}
are their traces with respect to the auxiliary space. Using \eqref{rh} (and accounting for ${\cal C}=2$), one may readily prove \cite{1,3,5} that ($I\equiv I^{(2^N)}$)
\begin{equation}\label{ttt}
t(\lambda)=U_L(I+2\lambda\hat H)+o(\lambda),\qquad
\tilde t(\lambda)=(I+2\lambda\hat H)U_R+o(\lambda),
\end{equation}
where
\begin{equation}
U_L={\rm tr}_aP_{1,a}\dots P_{N,a},\qquad U_R={\rm tr}_aP_{N,a}\dots P_{1,a},
\end{equation}
are the left and right shift operators. Namely, for $\xi_j\in{\mathbb C}^2$ ($j=1,\dots,N$)
\begin{equation}
U_L\xi_1\otimes\xi_2\otimes\dots\otimes\xi_N=\xi_2\otimes\xi_3\otimes\dots\otimes\xi_1,\qquad
U_R\xi_1\otimes\xi_2\otimes\dots\otimes\xi_N=\xi_N\otimes\xi_1\otimes\dots\otimes\xi_{N-1},
\end{equation}
so that
\begin{equation}\label{UU}
U_RU_L=U_LU_R=I.
\end{equation}

According to \eqref{ttt} and \eqref{UU},
\begin{equation}\label{b=a-aa}
\tilde t(-\nu)t(-\nu)=I-\frac{\beta\hat H}{\tt N}+o\Big(\frac{1}{\tt N}\Big),\qquad
\nu\equiv\frac{\beta}{4\tt N},
\end{equation}
where the parameter $\tt N$ is called the Trotter number \cite{2,3,4,5,6,7}. From \eqref{b=a-aa} and the Trotter-Suzuki formula \cite{2,3,5,7} follow that
\begin{equation}\label{expbH}
\lim_{\tt N\rightarrow\infty}[\tilde t(-\nu)t(-\nu)]^{\tt N}=
\lim_{\tt N\rightarrow\infty}\Big(I-\frac{\beta\hat H}{\tt N}\Big)^{\tt N}={\rm e}^{-\beta\hat H}.
\end{equation}
or, according to \eqref{t=TrT},
\begin{equation}\label{expbH2}
{\rm e}^{-\beta\hat H}=\lim_{{\tt N}\rightarrow\infty}{\rm tr}_{1,\dots,2{\tt N}}
\Big(\tilde T_1(-\nu)T_2(-\nu)\dots\tilde T_{2{\tt N}-1}(-\nu)T_{2{\tt N}}(-\nu)\Big).
\end{equation}

In itself, this formula is useless for the future treatment, for example, because (for example) due to the noncommutativity $[\tilde L_{N,1},\tilde L_{j,1}]\neq0$ and
$[L_{j,2},L_{N,2}]\neq0$ ($j\neq N$), the operators $\tilde L_{N,1}$ and $L_{N,2}$ in the product
\begin{equation}
\tilde T_1T_2=\tilde L_{N,1}\dots\tilde L_{1,1}L_{1,2}\dots L_{N,2},
\end{equation}
cannot been transferred to the neighboring positions. This lack however may be got over by the following trick.
Accounting for the invariance of trace under transposition (${\rm tr}A={\rm tr}A^t$), one may rewrite \eqref{expbH2} replacing
the matrices $\tilde T_j(-\nu)$ by their transposed with respect to auxiliary space
\begin{equation}
\tilde T_j(-\nu)\longrightarrow\tilde T_j^{t_2}(-\nu)=
\tilde L_{1,j}^{t_2}(-\nu)\dots\tilde L_{N,j}^{t_2}(-\nu).
\end{equation}
where $t_2$ means transposition in the second (auxiliary) space. Under this transposition-trick, \eqref{expbH2} turns into
\begin{equation}\label{expbH3}
{\rm e}^{-\beta\hat H}=\lim_{{\tt N}\rightarrow\infty}{\rm tr}_{1,\dots,2{\tt N}}T^{\rm QTM}_1(\lambda,\nu,{\tt N})T^{\rm QTM}_2(\lambda,\nu,
{\tt N})\dots T^{\rm QTM}_N(\lambda,\nu,{\tt N})|_{\lambda=0},
\end{equation}
where for $j=1,\dots,N$
\begin{equation}\label{TQTM}
T^{\rm QTM}_j(\lambda,\nu,{\tt N})=\tilde L_{j,1}^{t_2}(-\nu-\lambda)L_{j,2}(\lambda-\nu)\dots
\tilde L_{j,2{\tt N}-1}^{t_2}(-\nu-\lambda)L_{j,2{\tt N}}(\lambda-\nu),
\end{equation}
or equivalently
\begin{equation}\label{TQTM2}
T^{\rm QTM}_j(\lambda,\nu,{\tt N})=L^{\rm QTM}_{j(12)}(\lambda,\nu)L^{\rm QTM}_{j(34)}(\lambda,\nu)
\dots L^{\rm QTM}_{j(2{\tt N}-1\,2{\tt N})}(\lambda,\nu),
\end{equation}
where
\begin{equation}\label{lqtm}
L^{\rm QTM}_{j({\tt ab})}(\lambda,\nu)\equiv\tilde L_{j\tt a}^{t_2}(-\nu-\lambda)L_{j\tt b}(\lambda-\nu).
\end{equation}
According to \eqref{L} (and the identity $P^2=I^{(4)}$),
\begin{equation}\label{LtildeL}
\tilde L(\lambda)=PL(\lambda)P\Longleftrightarrow\tilde L_{ij}(\lambda)=L_{ji}(\lambda),
\end{equation}
so that \eqref{lqtm} may be represented in the equivalent form
\begin{equation}\label{lqtmstand}
L^{\rm QTM}_{j({\tt ab})}(\lambda,\nu)\equiv L_{{\tt a}j}^{t_1}(-\nu-\lambda)L_{j\tt b}(\lambda-\nu),
\end{equation}
adopted in \cite{4,5,6,7}.

Though the equivalence between \eqref{expbH2} and \eqref{expbH3} is rather elementary, it needs some comments.
All the $2{\tt N}$ factors inside the trace in the right side of \eqref{expbH2} are $2^N\times2^N$-matrices in the quantum space, whose
entries are $2\times2$ matrices in the corresponding copy of the $2{\tt N}$ auxiliary spaces. At the same time, each factor inside the trace in the right side of \eqref{expbH3} is $2\times2$ matrix in the corresponding local quantum space whose entries are $4^{\tt N}\times4^{\tt N}$-matrices in the so called Trotter space. The latter is the tensor product of all $2{\tt N}$ auxiliary spaces.

The main advantage of the representation \eqref{expbH3} is the permutation relation between the QTM $L$-operators
\begin{equation}\label{RLLQTM}
R_{12}(\lambda-\mu)L_{1(34)}^{\rm QTM}(\lambda,\nu)L_{2(34)}^{\rm QTM}(\mu,\nu)=L_{1(34)}^{\rm QTM}(\mu,\nu)L_{2(34)}^{\rm QTM}(\lambda,\nu)R_{12}(\lambda-\mu).
\end{equation}
It is similar to \eqref{RLL1} and, according to the definition \eqref{lqtm}, directly follows from \eqref{RLL1}, and \eqref{RLL2}, if the latter is
represented in an equivalent form
\begin{equation}\label{RLL3}
R_{12}(\lambda-\mu)\tilde L_{13}^{t_2}(-\nu-\lambda)\tilde L_{23}^{t_2}(-\nu-\mu)=\tilde L_{13}^{t_2}(-\nu-\mu)
\tilde L_{23}^{t_2}(-\nu-\lambda)R_{12}(\lambda-\mu).
\end{equation}
Following \eqref{RLLQTM} and \eqref{TQTM2},
\begin{equation}\label{RTTQTM}
R_{12}(\lambda-\mu)T_1^{\rm QTM}(\lambda,\nu,{\tt N})T_2^{\rm QTM}(\mu,\nu,{\tt N})=T_1^{\rm QTM}(\mu,\nu,{\tt N})T_2^{\rm QTM}(\lambda,\nu,{\tt N})R_{12}(\lambda-\mu).
\end{equation}

Taking the partition function as the trace in the quantum space
\begin{equation}
Z(\beta,N)={\rm Sp}_{1,\dots,N}{\rm e}^{-\beta\hat H},
\end{equation}
and postulating the ${\tt N}\rightarrow\infty$ interchangeability of the traces \cite{2,3,4,5,6,7}
\begin{equation}\label{Sptr=trSp}
{\rm Sp}_{1,\dots,N}\lim_{\tt N\rightarrow\infty}{\rm tr}_{1,\dots,2{\tt N}}=\lim_{\tt N\rightarrow\infty}{\rm tr}_{1,\dots,2{\tt N}}
{\rm Sp}_{1,\dots,N},
\end{equation}
one readily gets from \eqref{expbH3}
\begin{equation}\label{Z}
Z(\beta,N)=\lim_{{\tt N}\rightarrow\infty}{\rm tr}_{1,\dots,2{\tt N}}\Big(t^{\rm QTM}(0,\nu,{\tt N})\Big)^N,
\end{equation}
where
\begin{equation}\label{tQTM}
t^{\rm QTM}(\lambda,\nu,{\tt N})\equiv{\rm Sp}_jT^{\rm QTM}_j(\lambda,\nu,{\tt N}),
\end{equation}
is the $4^{\tt N}\times4^{\tt N}$ matrix in the Trotter space.

According to \eqref{Z} and \eqref{b=a-aa} the free energy density of the chain
\begin{equation}
f(\beta)\equiv-\frac{1}{\beta}\lim_{N\rightarrow\infty}\frac{1}{N}\ln{Z(\beta,N)},
\end{equation}
is
\begin{equation}\label{fQTMtr}
f(\beta)=-\frac{1}{\beta}\lim_{N\rightarrow\infty}\frac{1}{N}\ln{\lim_{\tt N\rightarrow\infty}
{\rm tr}_{\tt1,\dots,{\tt 2N}}\Big(t^{\rm QTM}(0,\nu,{\tt N})\Big)^N},\qquad\nu=\frac{\beta}{4\tt N},
\end{equation}
or in the expanded form
\begin{equation}\label{fQTMtr2}
f(\beta)=-\frac{1}{\beta}\lim_{N\rightarrow\infty}\frac{1}{N}\ln{\lim_{\tt N\rightarrow\infty}\sum_{k=1}^{4^{\tt N}}\Lambda_k^N(0,\beta,{\tt N})},
\end{equation}
where $\Lambda_k(\lambda,\beta,{\tt N})$ are the eigenvalues of $t^{\rm QTM}(\lambda,\nu,{\tt N})$.

According to \eqref{RTTQTM} and \eqref{tQTM}
\begin{equation}
[t^{\rm QTM}(\lambda,\nu,{\tt N}),t^{\rm QTM}(\mu,\nu,{\tt N})]=0.
\end{equation}
Hence the eigenvectors of $t^{\rm QTM}(\lambda,\nu,{\tt N})$ do not depend on $\lambda$.

The QTM machinery works if and only if
the matrix $t^{\rm QTM}(\lambda,\nu,{\tt N})$ has a {\it dominant} eigenvalue $\Lambda_{\rm max}(\lambda,\nu,{\tt N})$ \cite{2,3,4,5,6,7}.
This means that $\Lambda_{\rm max}(0,\nu,{\tt N})$ is the simple maximum (and, of course, positive) eigenvalue of $t^{\rm QTM}(0,\nu,{\tt N})$.
The corresponding ($\lambda$-independent) eigenvector $|V_{\rm max}(\nu,{\tt N})\rangle$, is called the {\it dominant} eigenvector
\begin{equation}\label{tV=LV}
t^{\rm QTM}(\lambda,\nu,{\tt N})|V_{\rm max}(\nu,{\tt N})\rangle=\Lambda_{\rm max}(\lambda,\nu,{\tt N})|V_{\rm max}(\nu,{\tt N})\rangle.
\end{equation}
Taking $\Lambda_1(\lambda,\nu,{\tt N})=\Lambda_{\rm max}(\lambda,\nu,{\tt N})$ one may represent \eqref{fQTMtr2} in the equivalent form
\begin{equation}\label{fQTMtr3}
f(\beta)=-\frac{1}{\beta}\lim_{N\rightarrow\infty}\frac{1}{N}\lim_{\tt N\rightarrow\infty}\Big[
N\ln{\Lambda_{\rm max}}(0,\nu,{\tt N})+\ln{\Big(1+\sum_{k=2}^{4^{\tt N}}
\Big(\frac{\Lambda_k(0,\nu,{\tt N})}{\Lambda_{\rm max}(0,\nu,{\tt N})}\Big)^{N}\Big)}\Big].
\end{equation}
So, under the condition
\begin{equation}\label{condtrL}
\lim_{N\rightarrow\infty}\frac{1}{N}\lim_{\tt N\rightarrow\infty}\ln{\Big[1+\sum_{k=2}^{4^{\tt N}}
\Big(\frac{\Lambda_k(0,\nu,{\tt N})}{\Lambda_{\rm max}(0,\nu,{\tt N})}\Big)^{N}\Big]}=0,
\end{equation}
the formula \eqref{fQTMtr2} reduces to \cite{3,4,5,6,7}
\begin{equation}\label{fQTM}
f(\beta)=-\frac{1}{\beta}\ln{\Lambda_{\infty}(0,\beta)},
\end{equation}
where
\begin{equation}\label{Laminfty}
\Lambda_{\infty}(\lambda,\beta)=\lim_{\tt N\rightarrow\infty}\Lambda_{\rm max}\Big(\lambda,\frac{\beta}{4\tt N},{\tt N}\Big).
\end{equation}
At the first glance, the information about $\Lambda_{\rm max}(\lambda,\nu,{\tt N})$ at $\lambda\neq0$ is unnecessary.
However, it will be employed in Sect. 6 and Sect. 7.

\section{Dominant eigenvalue at infinite temperature}

Following \eqref{rmat}, \eqref{L}, and \eqref{lqtm}
\begin{equation}\label{LQTMexpl}
L^{\rm QTM}(\lambda,\nu)=\left(\begin{array}{cc}
A(\lambda,\nu)&B(\lambda,\nu)\\
C(\lambda,\nu)&D(\lambda,\nu)
\end{array}\right),
\end{equation}
where
\begin{eqnarray}\label{LQTMentr}
&&A(\lambda,\nu)=\frac{1}{{\tt c}_+{\tt c}_-}\left(\begin{array}{cccc}
{\tt c}_+{\tt c}_-&0&0&1\\
0&{\tt c}_+{\tt s}_-&0&0\\
0&0&-{\tt s}_+{\tt c}_-&0\\
0&0&0&-{\tt s}_+{\tt s}_-
\end{array}\right),\nonumber\\
&&B(\lambda,\nu)=\frac{1}{{\tt c}_+{\tt c}_-}\left(\begin{array}{cccc}
0&0&{\tt s}_-&0\\
{\tt c}_+&0&0&{\tt c}_-\\
0&0&0&0\\
0&0&-{\tt s}_+&0
\end{array}\right),\qquad
C(\lambda,\nu)=\frac{1}{{\tt c}_+{\tt c}_-}\left(\begin{array}{cccc}
0&-{\tt s}_+&0&0\\
0&0&0&0\\
{\tt c}_-&0&0&{\tt c}_+\\
0&{\tt s}_-&0&0
\end{array}\right),\nonumber\\
&&D(\lambda,\nu)=\frac{1}{{\tt c}_+{\tt c}_-}\left(\begin{array}{cccc}
-{\tt s}_+{\tt s}_-&0&0&0\\
0&-{\tt s}_+{\tt c}_-&0&0\\
0&0&{\tt c}_+{\tt s}_-&0\\
1&0&0&{\tt c}_+{\tt c}_-
\end{array}\right),
\end{eqnarray}
and
\begin{equation}
{\tt s}_{\pm}\equiv\sin{(\lambda\pm\nu)},\qquad{\tt c}_{\pm}\equiv\cos{(\lambda\pm\nu)}.
\end{equation}

The substitution of \eqref{LQTMentr} into \eqref{LQTMexpl} yields
\begin{equation}\label{l0}
L^{\rm QTM}(0,0)=
\left(\begin{array}{cc}
|v_{11}\rangle\langle u|&|v_{12}\rangle\langle u|\\
|v_{21}\rangle\langle u|&|v_{22}\rangle\langle u|
\end{array}\right)=
M\otimes\langle u|,
\end{equation}
where
\begin{eqnarray}\label{vij}
&&|v_{11}\rangle=\left(\begin{array}{c}
1\\
0\\
0\\
0
\end{array}\right),
\qquad
|v_{12}\rangle=\left(\begin{array}{c}
0\\
1\\
0\\
0
\end{array}\right),
\qquad
|v_{21}\rangle=\left(\begin{array}{c}
0\\
0\\
1\\
0
\end{array}\right),
\qquad
|v_{22}\rangle=\left(\begin{array}{c}
0\\
0\\
0\\
1
\end{array}\right),\nonumber\\
&&\langle u|=\left(\begin{array}{cccc}
1&0&0&1
\end{array}\right),
\end{eqnarray}
and
\begin{equation}
M=\left(\begin{array}{cc}
|v_{11}\rangle&|v_{12}\rangle\\
|v_{21}\rangle&|v_{22}\rangle
\end{array}\right).
\end{equation}

The substitution of \eqref{TQTM2} and \eqref{l0} into \eqref{tQTM} results in
\begin{equation}\label{tQTM0}
t^{\rm QTM}(0,0,{\tt N})=|V\rangle\langle U|,
\end{equation}
where the vector
\begin{equation}\label{VTinfty}
|V\rangle={\rm tr}_0M_{01}\dots M_{0\tt N},
\end{equation}
or in the expanded representation
\begin{equation}\label{VTinftyexp}
|V\rangle=\sum_{j_1,\dots,j_{{\tt N}-1}}|v_{j_1j_2}\rangle\otimes|v_{j_2j_3}\rangle\otimes\dots\otimes|v_{j_{{\tt N}-1}j_1}\rangle,
\end{equation}
has the matrix product form \cite{13}, while
\begin{equation}
\langle U|=\langle u|^{\otimes\tt N}\equiv\langle u|\otimes\dots\otimes\langle u|.
\end{equation}

Since $\langle u|v_{11}\rangle=\langle u|v_{22}\rangle=1$ and $\langle u|v_{12}\rangle=\langle u|v_{21}\rangle=0$, one has
\begin{equation}\label{UV}
\langle U|V\rangle=(\langle u|v_{11}\rangle^{\tt N}+\langle u|v_{11}\rangle^{\tt N})=2.
\end{equation}
Hence, the matrix $t^{\rm QTM}(0,0,{\tt N})$ in \eqref{tQTM0} has the single non-zero eigenvalue
\begin{equation}\label{condLambda}
\Lambda_{\rm max}(0,0,{\tt N})=2.
\end{equation}
corresponding to the dominant eigenvector \eqref{VTinftyexp}. The verification of \eqref{condtrL} is trivial.

According to this result, we may conclude, that for $\nu<\infty$ the matrix $t^{\rm QTM}(\lambda,\nu,{\tt N})$ also should have
the single dominant eigenvalue $\Lambda_{\rm max}(\lambda,\nu,{\tt N})$, identified by the condition \eqref{condLambda},
so that the supplementary condition \eqref{condtrL} is satisfied.

\section{Evaluation of $\Lambda_{\rm max}(\lambda,\nu,{\tt N})$ by the QTM machinery}

Let
\begin{equation}
\hat  Q^{\rm QTM}=\sum_{n=1}^{\tt N}Q^{\rm QTM}_n,
\end{equation}
where the corresponding $4\times4$ density matrix has the form
\begin{equation}\label{QQTM}
Q^{\rm QTM}=\left(\begin{array}{cccc}
0&0&0&0\\
0&1&0&0\\
0&0&-1&0\\
0&0&0&0
\end{array}\right).
\end{equation}

It may be readily checked by direct calculations, that
\begin{equation}
[I^{(2)}\otimes Q^{\rm QTM},L^{\rm QTM}(\lambda,\nu)]=[{\bf S}^z\otimes I^{(4)},L^{\rm QTM}(\lambda,\nu)].
\end{equation}
Hence, according to \eqref{TQTM2},
\begin{equation}\label{[Q,T]}
[I^{(2)}\otimes\hat Q^{\rm QTM},T^{\rm QTM}(\lambda,\nu,{\tt N})]=[{\bf S}^z\otimes I^{(4^{\tt N})},T^{\rm QTM}(\lambda,\nu,{\tt N})].
\end{equation}
Suggesting the representation
\begin{equation}\label{TQTMentr}
T^{\rm QTM}(\lambda,\nu,{\tt N})\equiv\left(\begin{array}{cc}
\hat A(\lambda,\nu,{\tt N})&\hat B(\lambda,\nu,{\tt N})\\
\hat C(\lambda,\nu,{\tt N})&\hat D(\lambda,\nu,{\tt N})
\end{array}\right),
\end{equation}
one readily gets from \eqref{[Q,T]}
\begin{eqnarray}\label{[Q,AD]}
&&[\hat Q^{\rm QTM},\hat A(\lambda,\nu,{\tt N})]=[\hat Q^{\rm QTM},\hat D(\lambda,\nu,{\tt N})]=0\Longrightarrow
[\hat Q^{\rm QTM},t^{\rm QTM}(\nu,{\tt N})]=0,\\\label{[Q,B]}
&&{[}\hat Q^{\rm QTM},\hat B(\lambda,\nu,{\tt N})]=\hat B(\lambda,\nu,{\tt N}).
\end{eqnarray}

According to \eqref{QQTM} and \eqref{vij},  $Q^{\rm QTM}|v_{11}\rangle=Q^{\rm QTM}|v_{22}\rangle=0$. Hence,
\begin{equation}\label{QV1}
\hat Q^{\rm QTM}|V\rangle=0.
\end{equation}

Since the spectrum of $\hat Q^{\rm QTM}$ is integer, both the vectors $|V\rangle=|V_{\rm max}(0,{\tt N})\rangle$ and $|V_{\rm max}(\nu,{\tt N})\rangle$ should lie in the same sector of $\hat Q^{\rm QTM}$, and, according to \eqref{QV1},
\begin{equation}\label{QV2}
\hat Q^{\rm QTM}|V_{\rm max}(\nu,{\tt N})\rangle=0.
\end{equation}

From now we shall study only the case of even $\tt N$, implying
\begin{equation}\label{N=2M}
{\tt N}=2{\tt M},
\end{equation}
(the case of odd ${\tt N}$ is slightly more complex).

Following \eqref{RTTQTM},
\begin{eqnarray}\label{perm}
&&\hat A(\lambda,\nu,{\tt N})\hat B(\mu,\nu,{\tt N})=\cot{(\mu-\lambda)}\hat B(\mu,\nu,{\tt N})\hat A(\lambda,\nu,{\tt N})\nonumber\\
&&+\frac{1}{\sin{(\lambda-\mu)}}\hat B(\lambda,\nu,{\tt N})\hat A(\mu,\nu,{\tt N}),\nonumber\\
&&\hat D(\lambda,\nu,{\tt N})\hat B(\mu,\nu,{\tt N})=\cot{(\lambda-\mu)}\hat B(\mu,\nu,{\tt N})\hat D(\lambda,\nu,{\tt N})\nonumber\\
&&+\frac{1}{\sin{(\mu-\lambda)}}\hat B(\lambda,\nu,{\tt N})\hat D(\mu,\nu,{\tt N}),\nonumber\\
&&\hat B(\lambda,\nu,{\tt N})\hat B(\mu,\nu,{\tt N})=\hat B(\mu,\nu,{\tt N})
\hat B(\lambda,\nu,{\tt N}).
\end{eqnarray}

Let
\begin{equation}
|\downarrow\rangle=\left(\begin{array}{c}
0\\
1
\end{array}\right),\qquad
|\uparrow\rangle=\left(\begin{array}{c}
1\\
0
\end{array}\right),\qquad
|\downarrow\uparrow\rangle=|\downarrow\rangle\otimes|\uparrow\rangle=\left(\begin{array}{c}
0\\
0\\
1\\
0
\end{array}\right),
\end{equation}
and $|\emptyset\rangle$ is the tensor product of ${\tt N}$ factors
\begin{equation}\label{vac}
|\emptyset\rangle=|\downarrow\uparrow\rangle_1\dots|\downarrow\uparrow\rangle_{\tt N}\equiv|\downarrow\uparrow\rangle\otimes\dots\otimes|\downarrow\uparrow\rangle.
\end{equation}

According to \eqref{LQTMentr},
\begin{eqnarray}\label{Lvac}
&&A(\lambda,\nu)|\downarrow\uparrow\rangle=-\tan{(\lambda+\nu)}|\downarrow\uparrow\rangle,\qquad
D(\lambda,\nu)|\downarrow\uparrow\rangle=\tan{(\lambda-\nu)}|\downarrow\uparrow\rangle,\nonumber\\
&&C(\lambda,\nu)|\downarrow\uparrow\rangle=0.
\end{eqnarray}
Hence, following \eqref{TQTM2}, \eqref{vac}, \eqref{N=2M} and \eqref{Lvac}
\begin{equation}
\hat A(\lambda,\nu,{\tt N})|\emptyset\rangle=a(\lambda,\nu,{\tt N})|\emptyset\rangle,\qquad
\hat D(\lambda,\nu,{\tt N})|\emptyset\rangle=d(\lambda,\nu,{\tt N})|\emptyset\rangle,
\end{equation}
where
\begin{equation}\label{adequiv}
a(\lambda,\nu,{\tt N})\equiv\tan^{\tt N}{(\lambda+\nu)},\qquad d(\lambda,\nu,{\tt N})\equiv\tan^{\tt N}{(\lambda-\nu)}.
\end{equation}
At the same time, one may readily check that $Q^{\rm QTM}|\downarrow\uparrow\rangle=-|\downarrow\uparrow\rangle$, so that
\begin{equation}\label{Qempty}
\hat Q^{\rm QTM}|\emptyset\rangle=-{\tt N}|\emptyset\rangle.
\end{equation}
According to \eqref{[Q,B]} and \eqref{Qempty}, the condition \eqref{QV2} will be automatically satisfied if we suggest the vector
$|V_{\rm max}(\nu,{\tt N})\rangle$ in the form
\begin{equation}\label{Vmax}
|V_{\rm max}(\nu,{\tt N})\rangle=\hat B(\mu_1,\nu,{\tt N})\dots\hat B(\mu_{\tt N},\nu,{\tt N})|\emptyset\rangle,
\end{equation}
where $\{\mu_1,\dots,\mu_{\tt N}\}$ is a set of complex numbers.

Treating the state \eqref{Vmax} within the Bethe Ansatz machinery (and accounting for \eqref{N=2M}), one readily gets
($\hat B(\mu)\equiv\hat B(\mu,\nu,{\tt N})$)
\begin{eqnarray}\label{ADV}
&&\hat A(\lambda,\nu,{\tt N})|V_{\rm max}(\nu,{\tt N})\rangle=\tan^{\tt N}{(\lambda+\nu)}\prod_{j=1}^{2\tt M}\cot{(\lambda-\mu_j)}
|V_{\rm max}(\nu,{\tt N})\rangle\nonumber\\
&&+\hat B(\lambda,\nu,{\tt N})\sum_{j=1}^{2\tt M}\sigma_ja(\mu_j,\nu,{\tt N})
\hat B(\mu_1)\dots\hat B(\mu_{j-1})\hat B(\mu_{j+1})\dots\hat B(\mu_{\tt N})|\emptyset\rangle,\nonumber\\
&&\hat D(\lambda,\nu,{\tt N})|V_{\rm max}(\nu,{\tt N})\rangle=\tan^{\tt N}{(\lambda-\nu)}\prod_{j=1}^{2\tt M}\cot{(\lambda-\mu_j)}
|V_{\rm max}(\nu,{\tt N})\rangle\nonumber\\
&&+\hat B(\lambda,\nu,{\tt N})\sum_{j=1}^{2\tt M}\sigma_jd(\mu_j,\nu,{\tt N})
\hat B(\mu_1)\dots\hat B(\mu_{j-1})\hat B(\mu_{j+1})\dots\hat B(\mu_{\tt N})|\emptyset\rangle,
\end{eqnarray}
where
\begin{equation}
\sigma_j\equiv\frac{1}{\sin{(\lambda-\mu_j)}}\prod_{l\neq j}\cot{(\mu_l-\mu_j)}.
\end{equation}

Following \eqref{tV=LV}, \eqref{Vmax} and \eqref{ADV}
\begin{equation}\label{Lmax0}
\Lambda_{\max}(\lambda,\nu,{\tt N})=\Phi(\lambda,\nu,{\tt N})\prod_{j=1}^{2\tt M}\cot{(\lambda-\mu_j)},
\end{equation}
where (see \eqref{adequiv})
\begin{equation}\label{Phi}
\Phi(\lambda,\nu,{\tt N})\equiv a(\lambda,\nu,{\tt N})+d(\lambda,\nu,{\tt N})=\tan^{\tt N}{(\lambda+\nu)}+\tan^{\tt N}{(\lambda-\nu)},
\end{equation}
while the numbers $\mu_j$ satisfy the system of Bethe equations
\begin{equation}\label{Bethe}
\Phi(\mu_j,\nu,{\tt N})=0,
\end{equation}
whose solution is
\begin{equation}\label{Betheexp}
\frac{\tan{(\mu_j-\nu)}}{\tan{(\mu_j+\nu)}}=\kappa_j,\qquad
\kappa_j={\rm e}^{(2j-1)i\pi/{\tt N}},\qquad j=1,\dots,{\tt N}.
\end{equation}

Using the identity
\begin{equation}
\frac{\tan{(x+y)}-\tan{(x-y)}}{\tan{(x+y)}+\tan{(x-y)}}=\frac{\sin{2y}}{\sin{2x}},
\end{equation}
one reduces \eqref{Betheexp} to
\begin{equation}\label{sinmu}
\sin{2\mu_j}=\frac{1+\kappa_j}{1-\kappa_j}\sin{2\nu}=i\cot\frac{(2j-1)\pi}{2{\tt N}}\sin{2\nu}.
\end{equation}

For given $\sin{2\mu_j}$ there are two possible values of $\cot{\mu_j}$
\begin{equation}\label{cotpm}
\cot{\mu_j^{(\pm)}}=\frac{1}{\sin{2\mu_j}}\pm\sqrt{\frac{1}{\sin^2{2\mu_j}}-1},
\end{equation}
or, following \eqref{sinmu},
\begin{equation}\label{sinmupm}
\cot{\mu_j^{(\pm)}}
=i\Big(-\frac{\tan{[(2j-1)\pi/(2{\tt N})]}}{\sin{2\nu}}\pm \sqrt{1+\frac{\tan^2{[(2j-1)\pi/(2{\tt N})]}}{\sin^2{2\nu}}}\Big).
\end{equation}
Obviously,
\begin{eqnarray}\label{mu+mu}
&&\cot{\mu_j^{(+)}}\cot{\mu_j^{(-)}}=1,\\\label{limits}
&&\lim_{\nu\rightarrow0}\tan{\mu_j^{(-)}}=\lim_{\nu\rightarrow0}\tan{\mu_{{\tt M}+j}^{(+)}}=0,\qquad j=1,\dots,{\tt M},\nonumber\\
&&\lim_{\nu\rightarrow0}\cot{\mu_j^{(+)}}=\lim_{\nu\rightarrow0}\cot{\mu_{{\tt M}+j}^{(-)}}=0,\qquad j=1,\dots,{\tt M},
\end{eqnarray}
and
\begin{equation}\label{N+1-j}
\cot{\mu_{{\tt N}+1-j}^{(\pm)}}=-\cot{\mu_j^{(\mp)}},
\qquad\sin{2\mu_{{\tt N}+1-j}}=-\sin{2\mu_j}.
\end{equation}

Following \eqref{sinmupm},
\begin{equation}\label{cotcot}
\cot{\mu_j^{(-)}}\cot{\mu_{2{\tt M}+1-j}^{(+)}}=\Big(\frac{\tan{[(2j-1)\pi/(2{\tt N})]}}{\sin{2\nu}}+\sqrt{1+\frac{\tan^2{[(2j-1)\pi/(2{\tt N})]}}{\sin^2{2\nu}}}\Big)^2.
\end{equation}
Basing on \eqref{cotcot}, one may suggest the explicit expressions for the parameters $\mu_j$ in \eqref{Vmax}, by taking
\begin{equation}\label{mu=mump}
\mu_j=\mu^{(-)}_j,\quad j=1,\dots,{\tt M},\qquad\mu_j=\mu^{(+)}_j,\quad j={\tt M}+1,\dots,{\tt N}.
\end{equation}

Really, substituting \eqref{mu=mump} into \eqref{Lmax0} and accounting for \eqref{N=2M}, one gets
\begin{equation}\label{Lmax1}
\Lambda_{\max}(0,\nu,{\tt N})=\frac{2\sin^{\tt N}{2\nu}}{2^{\tt N}\cos^{2\tt N}{\nu}}
\prod_{j=1}^{\tt M}\cot{\mu_j^{(-)}}\prod_{j={\tt M}+1}^{2\tt M}\cot{\mu_j^{(+)}},
\end{equation}
or, according to \eqref{cotcot},
\begin{equation}\label{Lmax2}
\Lambda_{\rm max}(0,\nu,{\tt N})=\frac{2}{2^{\tt N}\cos^{2\tt N}\nu}\prod_{j=1}^{\tt M}\Big(\tan{\frac{(2j-1)\pi}{4{\tt M}}}
+\sqrt{\tan^2{\frac{(2j-1)\pi}{4{\tt M}}+\sin^2{2\nu}}}\Big)^2.
\end{equation}

Using the identity $\tan{(\pi/2-x)}=\cot{x}$, one may reduce \eqref{Lmax2} to the form
\begin{equation}\label{L=prodK}
\Lambda_{\rm max}(0,\nu,{\tt N})=2\prod_{j=1}^{\tt M}K_j(\nu,{\tt N}),
\end{equation}
where
\begin{eqnarray}
&&K_j(\nu,{\tt N})=\frac{1}{4\cos^4{\nu}}\Big(\tan{\frac{(2j-1)\pi}{4{\tt M}}}
+\sqrt{\tan^2{\frac{(2j-1)\pi}{4{\tt M}}+\sin^2{2\nu}}}\Big)\nonumber\\
&&\Big(\cot{\frac{(2j-1)\pi}{4{\tt M}}}
+\sqrt{\cot^2{\frac{(2j-1)\pi}{4{\tt M}}+\sin^2{2\nu}}}\Big).
\end{eqnarray}
Since,
\begin{equation}
K_j(0,{\tt N})=1,\qquad j=1,\dots,{\tt M},
\end{equation}
the condition \eqref{condLambda} is satisfied for \eqref{L=prodK}.

\section{The Bethe roots and the hole-type roots}

Following \eqref{mu=mump} it is convenient to represent the vector \eqref{Vmax} in a more precise form
\begin{equation}\label{Vmax3}
|V_{\rm max}(\nu,{\tt N})\rangle=\hat B(\lambda_1,\nu,{\tt N})\dots\hat B(\lambda_{\tt N},\nu,{\tt N})|\emptyset\rangle,
\end{equation}
where the $\tt N$ parameters $\lambda_j$,  defined by
\begin{equation}\label{Bethe_roots}
\cot{\lambda_j}=\begin{cases}
\cot{\mu_j^{(-)}},&j=1,\dots,{\tt M},\\
\cot{\mu_j^{(+)}},&j={\tt M}+1,\dots,2{\tt M},
\end{cases}
\end{equation}
are called the {\it Bethe roots} \cite{4,5,6,7}. The rest $\tt N$ parameters $w_j$, for which
\begin{equation}\label{hole_roots}
\cot{w_j}=\begin{cases}
\cot{\mu_{2{\tt M}+1-j}^{(-)}},&j=1,\dots,{\tt M},\\
\cot{\mu_{2{\tt M}+1-j}^{(+)}},&j={\tt M}+1,\dots,2{\tt M},
\end{cases}
\end{equation}
are called the {\it hole-type roots} \cite{4,5,6,7}.
Such separation of the parameters $\mu_j^{(\pm)}$ on the Bethe and hole-type roots, just supplies the explicit form \eqref{Vmax3} of the vector $|V_{\rm max}(\nu,{\tt N})\rangle$. Of course, it will be different for another eigenvector of $t^{\rm QTM}(\lambda,\nu,{\tt N})$.

The identities $\cot{it}=-i\coth{t}$ and $\cot{(it+\pi/2)}=-i\tanh{t}$ yield
\begin{eqnarray}\label{condtan}
&&\pm i(-|y|-\sqrt{1+y^2})=\cot{(\pm ix)},\qquad x>0,\quad \coth{x}=|y|+\sqrt{1+y^2},\nonumber\\
&&\pm i(|y|-\sqrt{1+y^2})=\cot{(\pm ix+\pi/2)},\qquad x>0,\quad \tanh{x}=\sqrt{1+y^2}-|y|.
\end{eqnarray}
With the account for \eqref{condtan}, one readily gets from \eqref{sinmupm}, \eqref{N+1-j}, \eqref{Bethe_roots}, and \eqref{hole_roots},
\begin{equation}\label{Re}
{\rm Re}\lambda_j=0,\qquad{\rm Re}w_j=\frac{\pi}{2},
\end{equation}
and
\begin{equation}\label{Im}
{\rm Im}\lambda_j={\rm Im}w_j=\begin{cases}
\alpha_j,\qquad j=1,\dots,{\tt M},\\
-\alpha_{2{\tt M}+1-j},\qquad j={\tt M}+1,\dots,2{\tt M},
\end{cases}
\end{equation}
where
\begin{equation}\label{kappa}
\tanh{\alpha_j}=\sqrt{1+\frac{\tan^2{[(2j-1)\pi/(2{\tt N})]}}{\sin^2{2\nu}}}-\frac{\tan{[(2j-1)\pi/(2{\tt N})]}}{\sin{2\nu}}.
\end{equation}
According to \eqref{Re} and \eqref{Im}
\begin{equation}\label{w=lam_pi2}
w_j=\lambda_j+\frac{\pi}{2}.
\end{equation}

As it follows from \eqref{Re}, \eqref{Im} and \eqref{kappa}, for fixed $j$ one has
\begin{equation}\label{accum}
\lim_{\tt N\rightarrow\infty}\lambda_j=0,\qquad\lim_{{\tt N}\rightarrow\infty}w_j=\frac{\pi}{2}.
\end{equation}
This formula expresses the accumulation of Bethe and hole-type roots at $\tt N\rightarrow\infty$, postulated for the XXZ model on the base
of numerical calculations \cite{4,5}.

Following \eqref{mu=mump} and \eqref{Bethe_roots}, one should rewrite \eqref{Lmax0} in the more transparent form
\begin{equation}\label{Lmaxprod}
\Lambda_{\max}(\lambda,\nu,{\tt N})=\Phi(\lambda,\nu,{\tt N})\prod_{j=1}^{2\tt M}\cot{(\lambda-\lambda_j)}.
\end{equation}

According to its definition \eqref{Phi}, the function $\Phi(\lambda,\nu,{\tt N})$ is the ratio of two polynomials of degree $4\tt N$ with respect
to ${\rm e}^{i\lambda}$ and
\begin{equation}\label{limPhi}
\lim_{\lambda\rightarrow i\infty}\Phi(\lambda,\nu,{\tt N})=2.
\end{equation}
Hence, the combination of \eqref{Phi}, \eqref{Bethe} and \eqref{limPhi} yields
\begin{equation}\label{Philambdaw}
\Phi(\lambda,\nu,{\tt N})=\frac{2\prod_{j=1}^{\tt N}\sin{(\lambda-\lambda_j)}\sin{(\lambda-w_j)}}
{[\cos{(\lambda+\nu)}\cos{(\lambda-\nu)}]^{\tt N}}.
\end{equation}
Substituting \eqref{Philambdaw} into \eqref{Lmaxprod}, and accounting for the equality
\begin{equation}
\prod_{j=1}^{2\tt M}\cot{(\lambda-\lambda_j)}=\prod_{j=1}^{2\tt M}\frac{\sin{(\lambda-w_j)}}{\sin{(\lambda-\lambda_j)}}
\end{equation}
which directly follows from \eqref{Re} and \eqref{Im}, one readily gets the representation
\begin{equation}\label{Lmaxw}
\Lambda_{\max}(\lambda,\nu,{\tt N})=\frac{2\prod_{j=1}^{\tt N}\sin^2{(\lambda-w_j)}}
{[\cos{(\lambda+\nu)}\cos{(\lambda-\nu)}]^{\tt N}}.
\end{equation}

\section{Evaluation of $\Lambda_{\infty}(0,\beta)$ by manipulations with contour integrals}

We suggest the contour $\gamma$ as two parallel lines ${\rm Re}z=-\pi/4$ and ${\rm Re}z=\pi/4$ in the complex plane and
the dual contour $\tilde\gamma$ as two parallel lines ${\rm Re}z=\pi/4$ and ${\rm Re}z=3\pi/4$. If
\begin{equation}\label{condg}
g(z+\pi)=g(z),\qquad\lim_{z\rightarrow i\infty}[g(z)-g(-z)]=0,
\end{equation}
then, obviously,
\begin{equation}\label{gammatgamma}
\oint_{\gamma+\tilde\gamma}dzg(z)=0\Longrightarrow\oint_{\gamma}dzg(z)=-\oint_{\tilde\gamma}dzg(z).
\end{equation}

Let now
\begin{equation}\label{fraka}
{\mathfrak a}(\lambda,\nu,{\tt N})\equiv\frac{d(\lambda,\nu,{\tt N})}{a(\lambda,\nu,{\tt N})}=\Big(\frac{\tan{(\lambda-\nu)}}{\tan{(\lambda+\nu)}}\Big)^{\tt N},
\end{equation}
and (see \eqref{Phi})
\begin{equation}\label{frakA}
{\mathfrak A}(\lambda,\nu,{\tt N})\equiv1+{\mathfrak a}(\lambda,\nu,{\tt N})=\frac{\Phi(\lambda,\nu,{\tt N})}
{\tan^{\tt N}{(\lambda+\nu)}},
\end{equation}
or according to \eqref{Philambdaw}
\begin{equation}
{\mathfrak A}(\lambda,\nu,{\tt N})=\frac{2\prod_{j=1}^{\tt N}\sin{(\lambda-\lambda_j)}\sin{(\lambda-w_j)}}
{[\sin{(\lambda+\nu)}\cos{(\lambda-\nu)}]^{\tt N}}.
\end{equation}
Since all the Bethe roots \eqref{Bethe_roots} lie inside $\gamma$, while all the hole-type roots \eqref{hole_roots} inside $\tilde\gamma$,
one has for rather small $\nu$ (big $\tt N$) and $\lambda\approx0$
\begin{equation}\label{int1}
\frac{1}{4\pi i}\oint_{\gamma}dz\tan{(\lambda-z)}\ln'{[{\mathfrak A}(z,\nu,{\tt N})]}
=\sum_{j=1}^{\tt N}\tan{(\lambda-\lambda_j)}-{\tt N}\tan{(\lambda+\nu)},
\end{equation}
and
\begin{equation}\label{tildeint1}
\frac{1}{4\pi i}\oint_{\tilde\gamma}dz\cot{(\lambda-z)}\ln'{[{\mathfrak A}(z,\nu,{\tt N})]}=\sum_{j=1}^{\tt N}\cot{(\lambda-w_j)}-
{\tt N}\cot{(\lambda-\nu-\pi/2)}.
\end{equation}

According to \eqref{w=lam_pi2}, one may rewrite \eqref{int1} in the form
\begin{equation}\label{int2}
\frac{1}{4\pi i}\oint_{\gamma}dz\tan{(\lambda-z)}\ln'{[{\mathfrak A}(z,\nu,{\tt N})]}
=-\sum_{j=1}^{\tt N}\cot{(\lambda-w_j)}-{\tt N}\tan{(\lambda+\nu)},
\end{equation}
more similar to \eqref{tildeint1}.

Since, inside the contour $\gamma$ the function ${\mathfrak A}(z,\nu,{\tt N})$ has ${\tt N}$ simple zeroes and the single ${\tt N}$-th order pole
its logarithm is unique defined on $\gamma$. The same is obviously true for the contour $\tilde\gamma$. Hence, the integration of \eqref{int2} and \eqref{tildeint1} by parts, with the account for \eqref{gammatgamma} yields
\begin{eqnarray}
&&-\frac{1}{4\pi i}\oint_{\gamma}dz\frac{\partial}{\partial z}\Big(\tan{(\lambda-z)}+\cot{(\lambda-z)}\Big)\ln{[{\mathfrak A}(z,\nu,{\tt N})]}\nonumber\\
&&=-2\sum_{j=1}^{\tt N}\cot{(\lambda-w_j)}-{\tt N}\tan{(\lambda+\nu)}-{\tt N}\tan{(\lambda-\nu)}.
\end{eqnarray}
Integrating now this formula with respect to $\lambda$, one readily gets ($C$ is a number)
\begin{equation}\label{int}
\frac{1}{2\pi i}\oint\frac{\ln{{\mathfrak A}(z,\nu,{\tt N})}dz}{\sin{2(z-\lambda)}}=
\ln{\frac{\prod_{j=1}^{\tt N}\sin^2{(\lambda-w_j)}}{[\cos{(\lambda+\nu)}\cos{(\lambda-\nu)}]^{\tt N}}}+C.
\end{equation}

According to \eqref{frakA} and \eqref{fraka}, the left side of \eqref{int} turns to zero at $\lambda\rightarrow i\infty$. Applying this requirement to the right side one gets $C=0$.
Now, the comparison between \eqref{Lmaxw} and \eqref{int} yields at ${\tt N}\rightarrow\infty$
\begin{equation}\label{Lmaxcontinf}
\ln{\Lambda_{\infty}(\lambda,\beta)}=\frac{1}{\pi i}\oint\frac{\ln{{\mathfrak A}_{\infty}(z,\beta)}dz}{\sin{2(z-\lambda)}},
\end{equation}
where
\begin{equation}
{\mathfrak A}_{\infty}(z,\beta)\equiv1+{\mathfrak a}_{\infty}(z,\beta),
\end{equation}
and
\begin{equation}
{\mathfrak a}_{\infty}(z,\beta)\equiv\lim_{{\tt N}\rightarrow\infty}{\mathfrak a}\Big(z,\frac{\beta}{4\tt N},{\tt N}\Big).
\end{equation}

Following \eqref{fraka} and \eqref{b=a-aa},
\begin{equation}\label{ainfty}
{\mathfrak a}_{\infty}(z,\beta)=\lim_{{\tt N}\rightarrow\infty}
\left(\frac{1-\frac{\sin{2\nu}}{\sin{2z}}}{1+\frac{\sin{2\nu}}{\sin{2z}}}\right)^{\tt N}=
{\rm e}^{-\frac{\beta}{\sin{2z}}}.
\end{equation}
So, taking
\begin{equation}\label{contour}
z=ip\pm\frac{\pi}{4},\qquad p\in(-\infty,\infty),
\end{equation}
on the right and left sides of the contour $\gamma$ and substituting \eqref{ainfty} into \eqref{Lmaxcontinf} one gets
\begin{equation}\label{Lmaxlines}
\ln{\Lambda_{\infty}(0,\beta)}=\frac{1}{\pi}\int_{-\infty}^{\infty}\frac{dp}{\cosh{2p}}\Big[
\ln{\Big(1+{\rm e}^{-\frac{\beta}{\cosh{2p}}}\Big)}+\ln{\Big(1+{\rm e}^{\frac{\beta}{\cosh{2p}}}\Big)}\Big].
\end{equation}
Taking
\begin{equation}\label{substint}
\frac{1}{\cosh{2p}}=\cos{k},\qquad\tanh{2p}=\sin{k},\qquad dk=\frac{2dp}{\cosh{2p}},
\end{equation}
one reduces \eqref{Lmaxlines} to the canonical form
\begin{equation}\label{Lmaxcanon}
\ln{\Lambda_{\infty}(0,\beta)}=\frac{1}{2\pi}\int_{-\pi}^{\pi}dk\ln{\Big(1+{\rm e}^{-\beta\cos{k}}\Big)},
\end{equation}
which, according to \eqref{fQTM} gives the well known formula \cite{11}
\begin{equation}\label{fh0}
f(\beta)=-\frac{1}{2\pi\beta}\int_{-\pi}^{\pi}dk\ln{\Big(1+{\rm e}^{-\beta\cos{k}}\Big)}.
\end{equation}

\section{Evaluation of $\Lambda_{\infty}(0,\beta)$ by manipulations with Fourier transformations}

Following \cite{4,5,6,7}, we introduce the new variables
\begin{equation}
\bar{\mathfrak a}(\lambda,\nu,{\tt N})\equiv\frac{1}{{\mathfrak a}(\lambda,\nu,{\tt N})},\qquad
\bar{\mathfrak A}(\lambda,\nu,{\tt N})\equiv1+\bar{\mathfrak a}(\lambda,\nu,{\tt N}),
\end{equation}
dual to ${\mathfrak a}(\lambda,\nu,{\tt N})$ and ${\mathfrak A}(\lambda,\nu,{\tt N})$. According to \eqref{fraka} and \eqref{frakA},
\begin{equation}
\bar{\mathfrak a}(\lambda,\nu,{\tt N})={\mathfrak a}(\lambda+\pi/2,\nu,{\tt N}),\qquad
\bar{\mathfrak A}(\lambda,\nu,{\tt N})={\mathfrak A}(\lambda+\pi/2,\nu,{\tt N}).
\end{equation}

By analogy with \eqref{frakA}, one has
\begin{equation}\label{barfrakA}
\bar{\mathfrak A}(\lambda,\nu,{\tt N})=\frac{\Phi(\lambda,\nu,{\tt N})}{\tan^{\tt N}{(\lambda-\nu)}},
\end{equation}
Following \eqref{Lmaxprod}, \eqref{frakA} and \eqref{barfrakA}
\begin{equation}\label{LmaxfrakAA}
\Lambda_{\max}(\lambda,\nu,{\tt N})=\frac{{\mathfrak A}(\lambda,\nu,{\tt N})\prod_{j=1}^{2\tt M}\cot{(\lambda-\lambda_j)}}{\cot^{\tt N}{(\lambda+\nu)}}
=\frac{\bar{\mathfrak A}(\lambda,\nu,{\tt N})\prod_{j=1}^{2\tt M}\cot{(\lambda-\lambda_j)}}{\cot^{\tt N}{(\lambda-\nu)}},
\end{equation}
so that
\begin{equation}\label{LamLam}
\Lambda_{\max}\Big(\lambda,\nu,{\tt N}\Big)\Lambda_{\max}\Big(\lambda+\frac{\pi}{2},\nu,{\tt N}\Big)
={\mathfrak A}\Big(\lambda,\nu,{\tt N}\Big)\bar{\mathfrak A}\Big(\lambda+\frac{\pi}{2},\nu,{\tt N}\Big)\bar{\mathfrak a}\Big(\lambda,\nu,{\tt N}\Big).
\end{equation}

Taking the limit ${\tt N}\rightarrow\infty$, and accounting for \eqref{ainfty}, one readily gets from \eqref{LamLam}
\begin{equation}\label{ln+ln}
\ln{\Lambda_{\infty}\Big(\lambda,\beta\Big)}+\ln{\Lambda_{\infty}\Big(\lambda+\frac{\pi}{2},\beta\Big)}=
\ln{\Big(1+{\rm e}^{-\frac{\beta}{\sin{2\lambda}}}\Big)}+\ln{\Big(1+{\rm e}^{\frac{\beta}{\sin{2\lambda}}}\Big)},
\end{equation}
or equivalently
\begin{equation}\label{ln+ln2}
\ln{\Lambda_{\infty}\Big(\lambda-\frac{\pi}{4},\beta\Big)}+\ln{\Lambda_{\infty}\Big(\lambda+\frac{\pi}{4},\beta\Big)}=
\ln{\Big(1+{\rm e}^{-\frac{\beta}{\cos{2\lambda}}}\Big)}+\ln{\Big(1+{\rm e}^{\frac{\beta}{\cos{2\lambda}}}\Big)}.
\end{equation}

Let now
\begin{equation}\label{tLambda}
\tilde\Lambda_{\infty}(p,\beta)\equiv\Lambda_{\infty}(ip,\beta).
\end{equation}
Following \eqref{ln+ln2}
\begin{equation}\label{tln+ln}
\ln{\tilde\Lambda_{\infty}\Big(p-\frac{i\pi}{4},\beta\Big)}+\ln{\tilde\Lambda_{\infty}\Big(p+\frac{i\pi}{4},\beta\Big)}=
\ln{\Big(1+{\rm e}^{-\frac{\beta}{\cosh{2p}}}\Big)}+\ln{\Big(1+{\rm e}^{\frac{\beta}{\cosh{2p}}}\Big)}.
\end{equation}

Following \cite{4,5}, the equation
\begin{equation}\label{F+F=G}
F\Big(p-\frac{i\pi}{4}\Big)+F\Big(p+\frac{i\pi}{4}\Big)=G(p),
\end{equation}
may be solved by manipulations with Fourier transformations. Taking the notations
\begin{equation}\label{Fourierg}
g(p)=\int_{-\infty}^{\infty}dx{\rm e}^{ipx}\hat g(x),\qquad\hat g(x)\equiv\frac{1}{2\pi}\int_{-\infty}^{\infty}dp{\rm e}^{-ipx}g(p),
\end{equation}
and one readily gets from \eqref{F+F=G}
\begin{equation}
\hat F(x)=\frac{\hat G(x)}{{\rm e}^{\pi x/4}+{\rm e}^{-\pi x/4}}.
\end{equation}

Since
\begin{eqnarray}
&&\int_{-\infty}^{\infty}\frac{{\rm e}^{ipx}dx}{{\rm e}^{\pi x/4}+{\rm e}^{-\pi x/4}}=2\pi i\sum_{j=0}^{\infty}
{\frac{2{\rm e}^{-2(2j+1)p}}{\pi i\sin{[(2j+1)\pi/2]}}}\nonumber\\
&&=4{\rm e}^{-2p}\sum_{j=0}^{\infty}(-1)^j{\rm e}^{-4pj}=\frac{2}{\cosh{2p}},
\end{eqnarray}
one has
\begin{equation}
F(q)=\frac{1}{\pi}\int_{-\infty}^{\infty}\frac{G(p)dp}{\cosh{2(q-p)}}.
\end{equation}
Using this formula, and accounting for \eqref{tLambda} and \eqref{tln+ln} one readily gets \eqref{Lmaxlines}.

\section{Account of magnetic field}

In \cite{12} the XX model was considered in the presence of a magnetic field. The corresponding Hamiltonian is
\begin{equation}
\hat H(h)=\hat H-h\hat {\bf S}^z,
\end{equation}
where
\begin{equation}
\hat {\bf S}^z=\sum_{n=1}^N{\bf S}^z_n.
\end{equation}
Since $[\hat {\bf S}^z,\hat H]=0$, one has from \eqref{expbH3}

\begin{equation}\label{expbH3h}
{\rm e}^{-\beta\hat H(h)}={\rm e}^{\beta h\hat{\bf S}^z}\lim_{\tt N\rightarrow\infty}{\rm tr}_{\tt1,\dots,{\tt 2N}}T^{\rm QTM}_1(\lambda,\nu,{\tt N})T^{\rm QTM}_2(\lambda,\nu,
{\tt N})\dots T^{\rm QTM}_N(\lambda,\nu,{\tt N})|_{\lambda=0},
\end{equation}
or equivalently
\begin{equation}\label{expbHh}
{\rm e}^{-\beta\hat H(h)}=\lim_{\tt N\rightarrow\infty}{\rm tr}_{\tt1,\dots,{\tt 2N}}T^{\rm QTM}_1(\lambda,h,\nu,{\tt N})T^{\rm QTM}_2(\lambda,h,\nu,
{\tt N})\dots T^{\rm QTM}_N(\lambda,h,\nu,{\tt N})|_{\lambda=0},
\end{equation}
where
\begin{equation}
T^{\rm QTM}(\lambda,h,\nu,{\tt N})=({\rm e}^{\beta h{\bf S}^z}\otimes I^{(4^{\tt N})})T^{\rm QTM}(\lambda,\nu,{\tt N}).
\end{equation}
In other words, $T^{\rm QTM}(\lambda,h,\nu,{\tt N})$ has the form \eqref{TQTMentr}, however with
\begin{eqnarray}
\hat A^{\rm QTM}(\lambda,h,\nu,{\tt N})={\rm e}^{\beta h/2}\hat A^{\rm QTM}(\lambda,\nu,{\tt N}),\quad
\hat B^{\rm QTM}(\lambda,h,\nu,{\tt N})={\rm e}^{\beta h/2}\hat B^{\rm QTM}(\lambda,\nu,{\tt N}),\nonumber\\
\hat C^{\rm QTM}(\lambda,h,\nu,{\tt N})={\rm e}^{-\beta h/2}\hat C^{\rm QTM}(\lambda,\nu,{\tt N}),\quad
\hat D^{\rm QTM}(\lambda,h,\nu,{\tt N})={\rm e}^{-\beta h/2}\hat D^{\rm QTM}(\lambda,\nu,{\tt N}).
\end{eqnarray}
It may be readily checked, that the relations \eqref{perm} are invariant under the substitution
$T^{\rm QTM}(\lambda,\nu,{\tt N})\rightarrow T^{\rm QTM}(\lambda,h,\nu,{\tt N})$. As the result (disregarding the factor ${\rm e}^{\beta h{\tt N}/2}$)
one may put $|V_{\rm max}(h,\nu,{\tt N})\rangle=|V_{\rm max}(\nu,{\tt N})\rangle$.
The system \eqref{ADV} will turn into
\begin{eqnarray}\label{ADVh}
&&\hat A(\lambda,h,\nu,{\tt N})|V_{\rm max}(\nu,{\tt N})\rangle={\rm e}^{\beta h/2}\tan^{\tt N}{(\lambda+\nu)}\prod_{j=1}^{2\tt M}\cot{(\lambda-\mu_j)}
|V_{\rm max}(\nu,{\tt N})\rangle+\dots,\nonumber\\
&&\hat D(\lambda,h,\nu,{\tt N})|V_{\rm max}(\nu,{\tt N})\rangle={\rm e}^{-\beta h/2}\tan^{\tt N}{(\lambda-\nu)}\prod_{j=1}^{2\tt M}\cot{(\lambda-\mu_j)}
|V_{\rm max}(\nu,{\tt N})\rangle+\dots.\qquad
\end{eqnarray}
Correspondingly, \eqref{adequiv} should be replaced by
\begin{equation}\label{adequivh}
a(\lambda,\nu,h,{\tt N})\equiv{\rm e}^{\beta h/2}\tan^{\tt N}{(\lambda+\nu)},\qquad d(\lambda,\nu,h,{\tt N})\equiv
{\rm e}^{-\beta h/2}\tan^{\tt N}{(\lambda-\nu)}.
\end{equation}
The dominant eigenvalue \eqref{Lmax0} and the system of Bethe equations \eqref{Betheexp} will take the forms
\begin{equation}\label{Lmaxh}
\Lambda_{\max}(h,\nu,{\tt N})=[{\rm e}^{\beta h/2}\tan^{\tt N}{(\lambda+\nu)}+{\rm e}^{-\beta h/2}\tan^{\tt N}{(\lambda-\nu)}]
\prod_{j=1}^{2\tt M}\cot{[\lambda-\mu_j(h)]},
\end{equation}
and (see \eqref{b=a-aa})
\begin{equation}\label{Betheexph}
\frac{\tan{(\mu_j(h)-\nu)}}{\tan{(\mu_j(h)+\nu)}}={\rm e}^{4\nu h}\kappa_j,\qquad
\kappa_j={\rm e}^{(2j-1)i\pi/{\tt N}},\qquad j=1,\dots,{\tt N}.
\end{equation}
The representation \eqref{sinmupm} should be replaced by
\begin{equation}\label{coth}
\cot{\mu_j^{(\pm)}(h)}
=i\Big(-\frac{\tan{\theta_j(h)}}{\sin{2\nu}}\pm \sqrt{1+\frac{\tan^2{\theta_j(h)}}{\sin^2{2\nu}}}\Big),
\end{equation}
where
\begin{equation}\label{theta}
\theta_j(h)\equiv\frac{(2j-1)\pi}{2{\tt N}}-2i\nu h.
\end{equation}
The separation on Bethe and hole-type roots will be the same as in \eqref{Bethe_roots} and \eqref{hole_roots}.
The formulas \eqref{Lmaxw} and \eqref{fraka} will turn into
\begin{equation}\label{Lmaxwh}
\Lambda_{\max}(\lambda,h,\nu,{\tt N})=\frac{({\rm e}^{\beta h/2}+{\rm e}^{-\beta h/2})\prod_{j=1}^{\tt N}\sin^2{(\lambda-w_j)}}
{[\cos{(\lambda+\nu)}\cos{(\lambda-\nu)}]^{\tt N}},
\end{equation}
and
\begin{equation}\label{frakah}
{\mathfrak a}(z,h,\nu,{\tt N})\equiv{\rm e}^{-\beta h}\Big(\frac{\tan{(z-\nu)}}{\tan{(z+\nu)}}\Big)^{\tt N}.
\end{equation}

As the result, instead of \eqref{Lmaxcontinf}, there should be
\begin{equation}\label{Lmaxcontinfh}
\ln{\Lambda_{\infty}(\lambda,h,\beta)}=\ln{\Big(\frac{{\rm e}^{\beta h/2}+{\rm e}^{-\beta h/2}}{1+{\rm e}^{-\beta h}}\Big)}+\frac{1}{\pi i}\oint_{\gamma}\frac{\ln{\mathfrak A}_{\infty}(z,h,\beta)}{\sin{2(z-\lambda)}}dz,
\end{equation}
where
\begin{equation}
{\mathfrak A}_{\infty}(z,h,\beta)=1+{\rm e}^{-\beta[h+1/\sin(2z)]}.
\end{equation}
The substitution \eqref{substint} yields
\begin{equation}\label{Lmaxcanon}
\ln{\Lambda_{\infty}(0,h,\beta)}=\frac{\beta h}{2}+\frac{1}{2\pi}\int_{-\pi}^{\pi}dk\ln{\Big(1+{\rm e}^{-\beta(h+\cos{k})}\Big)},
\end{equation}
which, according to \eqref{fQTM}, results in the analog of \eqref{fh0}
\begin{equation}\label{fhfin}
f(\beta,h)=-\frac{h}{2}-\frac{1}{2\pi\beta}\int_{-\pi}^{\pi}dk\ln{\Big(1+{\rm e}^{-\beta(h+\cos{k})}\Big)}.
\end{equation}

Using the auxiliary formula
\begin{equation}
\ln{\Big(1+{\rm e}^{-\beta(h+\cos{k})}\Big)}=-\frac{\beta(h+\cos{k})}{2}+
\ln{\Big({\rm e}^{\beta(h+\cos{k})/2}+{\rm e}^{-\beta(h+\cos{k})/2}\Big)},
\end{equation}
one may reduce \eqref{fhfin} to the well known expression \cite{12}
\begin{equation}\label{fhkats}
f(\beta,h)=-\frac{1}{\pi\beta}\int_0^{\pi}dk\ln{\Big(2\cosh{\frac{\beta(h+\cos{k})}{2}}\Big)}.
\end{equation}

\section{Summary and conclusions}

In the present paper, basing on the QTM approach, we gave the detailed and self-consistent derivation for the free energy density of the XX spin chain in zero magnetic field and briefly explained the modifications necessary at non-zero field. The resulting formula \eqref{fhkats}
(the integral representation for the free energy density at a non-zero magnetic field) has been obtained long ago \cite{12}, but within the alternative approach. It was also derived in \cite{2} on the base of the Trotter formula, however without use of the QTM machinery (based on \eqref{YB} and
\eqref{RTTQTM}).

The QTM formula for the free energy density of the (more general than XX) XXZ model also has been previously given in
the fundamental QTM texts \cite{4,5,6,7}. However, the result was not presented with full clarity and the derivation contained some gaps.
In the present paper treating the XX model (the special reduction of XXZ the one) we have filled some of these gaps. Namely:
We obtained the simple matrix-product representation \eqref{VTinfty} for the dominant eigenvector
at infinite temperature. For finite temperatures we rigorously (without any references on numerical calculations) proved for it
the rather simple {\it analytical} formula \eqref{Vmax3}. At zero magnetic field we derived the exact representations for the Bethe \eqref{Bethe_roots} and hole-type roots \eqref{hole_roots}. On the whole, we have shown that the QTM algebraic machinery, in itself, is rather elementary and clear. At the same time, we do not discuss the validity of formulas \eqref{Sptr=trSp} and \eqref{condtrL}
(the interested reader may turn to \cite{2} and \cite{6}).

The additional complexities, presented in \cite{4,5,6,7}, are not inherent in the QTM approach, but originate from the complexity of the XXZ model
for which the explicit representations for the finite-$N$ wave functions are absent.
Namely, in this case the existence of dominant eigenvector and its representation (similar to \eqref{Vmax3}) are postulated basing on the (not published) results of numerical experiments. Both the Bethe, and the hole-type roots were not presented explicitly in \cite{4,5,6,7}. As a result, the principal difference between them, as well as their accumulations in the ${\tt N}\rightarrow\infty$ limit \eqref{accum} may be rather unclear for an inexperienced reader. In the XXZ case the function
${\mathfrak a}(\lambda)$ does not have the simple explicit form, similar to \eqref{fraka}, but satisfy the integral equation, whose analytic solution is known only in the Ising case \cite{8} (and may be perturbatively studied at its vicinity \cite{14}).

The paper has the double task. From one side, it emphasizes the basic constructions inherent just in the QTM approach. From the other, it gives the
maximally detailed, step by step description of the calculation machinery.
We believe that the paper will be useful for beginners and specialists in adjacent areas.

The author is grateful to Hermann Boos, Frank ${\rm G\ddot ohmann}$, Andreas ${\rm Kl\ddot umper}$, Karol Kozlowski, and Sergei Rutkevich for the helpful discussions.

\end{document}